\documentclass[letterpaper]{article} % DO NOT CHANGE THIS
\usepackage{aaai24}
\usepackage{times}  % DO NOT CHANGE THIS
\usepackage{helvet}  % DO NOT CHANGE THIS
\usepackage{courier}  % DO NOT CHANGE THIS
\usepackage[hyphens]{url}  % DO NOT CHANGE THIS
\usepackage{graphicx} % DO NOT CHANGE THIS
\usepackage{microtype}

\usepackage{tabularx}
\usepackage{amsmath}

\usepackage{mathrsfs}
\usepackage{amsfonts}
\usepackage{dutchcal}
\usepackage{multirow}
\usepackage{subfigure}

\usepackage{amsmath,amsfonts}
\usepackage{array}
\usepackage{booktabs}
 
\usepackage[skip=1pt,labelfont=bf]{caption}
 \usepackage{calligra}
\usepackage{color}
\usepackage{colortbl}
\usepackage{courier}
\usepackage{csvsimple}
\usepackage{enumitem}
\usepackage{fancybox}
\usepackage{fontenc}
\usepackage{graphicx}

\usepackage{listings}
\usepackage{longtable}
\usepackage{lscape}
\usepackage{makecell}
\usepackage{marvosym}
\usepackage{moreverb}
\usepackage{multicol}
\usepackage{multirow}
\usepackage{makecell}
\usepackage{pifont}
\usepackage{rotating}

\usepackage[most]{tcolorbox}
\usepackage{threeparttable}
\usepackage{tikz}
\usepackage{soul}
\usepackage{url}
\usepackage{soul}
\usepackage{wasysym}
\usepackage{xspace}
\usepackage{amsthm}

\newcolumntype{L}[1]{>{\raggedright\let\newline\\\arraybackslash\hspace{0pt}}m{#1}}
\newcolumntype{C}[1]{>{\centering\let\newline\\\arraybackslash\hspace{0pt}}m{#1}}
\newcolumntype{R}[1]{>{\raggedleft\let\newline\\\arraybackslash\hspace{0pt}}m{#1}}

\definecolor{codegreen}{rgb}{0,0.6,0}
\definecolor{codered}{rgb}{1,0,0}
\definecolor{codegray}{rgb}{0.9,0.9,0.9}
\definecolor{codepurple}{rgb}{0.58,0,0.82}
\definecolor{backcolour}{rgb}{0.95,0.95,0.92}
\definecolor{lightgray}{gray}{0.9}

\lstdefinestyle{mystyle}{
    commentstyle=\color{codegreen},
    keywordstyle=\color{magenta},
    numberstyle=\small\color{black},
    stringstyle=\color{codepurple},
    basicstyle=\scriptsize\ttfamily,
    breakatwhitespace=false,
    breaklines=true,
    captionpos=b,
    keepspaces=true,
    showspaces=false,
    showstringspaces=false,
    showtabs=false,
    tabsize=2
}

\lstdefinelanguage{diff}{
  morecomment=[f][\color{blue}]{@@},     % group identifier
  morecomment=[f][\color{red}]-,         % deleted lines
  morecomment=[f][\color{codegreen}]+,       % added lines
  morecomment=[f][\color{red}]{---}, % Diff header lines (must appear after +,-)
  morecomment=[f][\color{codegreen}]{+++},
}

\setlist{noitemsep} %to leave space around whole list

\definecolor{darkpastelred}{rgb}{0.76, 0.23, 0.13}
\definecolor{ao(english)}{rgb}{0.0, 0.5, 0.0}

\definecolor{darkpastelred}{rgb}{0.76, 0.23, 0.13}
\definecolor{ao(english)}{rgb}{0.0, 0.5, 0.0}

\definecolor{yellow}{RGB}{255,255,153}
\definecolor{grey}{RGB}{224,224,224}

\newboolean{showcomments}
\setboolean{showcomments}{true}
\ifthenelse{\boolean{showcomments}}
 { \newcommand{\mynote}[2]{
      \fbox{\bfseries\sffamily\scriptsize#1}
        {\small$\blacktriangleright$\textsf{\emph{#2}}$\blacktriangleleft$}}}
        { \newcommand{\mynote}[2]{}}

\definecolor{DarkOrange}{rgb}{0.8,0.3,0.0}
\definecolor{DarkCyan}{rgb}{0.0, 0.55, 0.55}
\definecolor{DarkCyel}{rgb}{1.0, 0.49, 0.0}
\definecolor{yellow-green}{rgb}{0.6, 0.8, 0.2}

\newcolumntype{?}{!{\vrule width 1pt}}

\newcommand*{\ie}{i.e., }

\newcommand{\tool}{\texttt{HyCoQA}\xspace}

\newcommand{\find}[1]{
\begin{tcolorbox}[leftrule=0.2mm,toprule=0mm,bottomrule=0mm,left=0.0pt,right=0pt,top=0pt,bottom=0pt]%[tile,size=fbox,boxsep=2mm,boxrule=0pt,top=0pt,bottom=0pt,borderline={0.5mm}{0pt}{black!70!white},colback=black!5!white]
\em #1
\end{tcolorbox}
}

\newcommand{\equalcontrib}{\thanks{Equal contribution}}
\newcommand{\corrauthor}{\thanks{Corresponding author}}

\usepackage{natbib}  % DO NOT CHANGE THIS AND DO NOT ADD ANY OPTIONS TO IT
\usepackage{caption} % DO NOT CHANGE THIS AND DO NOT ADD ANY OPTIONS TO IT
\usepackage{algorithm}
\usepackage{algorithmic}

\usepackage{newfloat}
\usepackage{listings}

\DeclareCaptionStyle{ruled}{labelfont=normalfont,labelsep=colon,strut=off} % DO NOT CHANGE THIS
\floatstyle{ruled}
\newfloat{listing}{tb}{lst}{}
\floatname{listing}{Listing}
\title{Hyperbolic Code Retrieval: A Novel Approach for Efficient Code Search Using Hyperbolic Space Embeddings}
\author{
    Xunzhu Tang\textsuperscript{\rm 1}\corrauthor,
    Zhenghan Chen\textsuperscript{\rm 2}\equalcontrib,
    Saad Ezzini\textsuperscript{\rm 3},
    Haoye Tian\textsuperscript{\rm 1},
    Yewei Song\textsuperscript{\rm 1},
    Jacques KLEIN\textsuperscript{\rm 1},
    Tegawendé F. Bissyandé\textsuperscript{\rm 1}
}
\affiliations{
    \textsuperscript{\rm 1}University of Luxembourg, Luxembourg\\
    \textsuperscript{\rm 2}Peking University, Beijing, China\\
    \textsuperscript{\rm 3}Lancaster University, Lancaster, UK\\

    xunzhu.tang@uni.lu,
    1979282882@pku.edu.cn,
    s.ezzini@lancaster.ac.uk,
    haoye.tian@uni.lu,
    yewei.song@uni.lu,
    jacques.klein@uni.lu,
    tegawende.bissyande@uni.lu
}

\usepackage{bibentry}
\begin{document}

\maketitle

\begin{abstract}
Within the realm of advanced code retrieval, existing methods have primarily relied on intricate matching and attention-based mechanisms. However, these methods often lead to computational and memory inefficiencies, posing a significant challenge to their real-world applicability.
To tackle this challenge, we propose a novel approach, the Hyperbolic Code QA Matching (\tool). This approach leverages the unique properties of Hyperbolic space to express connections between code fragments and their corresponding queries, thereby obviating the necessity for intricate interaction layers. The process commences with a reimagining of the code retrieval challenge, framed within a question-answering (QA) matching framework, constructing a dataset with triple matches characterized as \texttt{<}negative code, description, positive code\texttt{>}. These matches are subsequently processed via a static BERT embedding layer, yielding initial embeddings. Thereafter, a hyperbolic embedder transforms these representations into hyperbolic space, calculating distances between the codes and descriptions. The process concludes by implementing a scoring layer on these distances and leveraging hinge loss for model training. Especially, the design of \tool{} inherently facilitates self-organization, allowing for the automatic detection of embedded hierarchical patterns during the learning phase. Experimentally, \tool{} showcases remarkable effectiveness in our evaluations: an average performance improvement of 3.5\% to 4\% compared to state-of-the-art code retrieval techniques.

\end{abstract}

\section{Introduction}
\label{sec:intro}
%Software development, as a discipline, has witnessed a transformation from being merely a technical endeavor to an intricate amalgamation of technology, semantics, and collaboration. The burgeoning proliferation of open-source platforms, such as GitHub and StackOverflow, has democratized access to a vast expanse of code repositories. Developers now, more than ever, are positioned in an ocean of information. %Yet, the challenge persists: one efficiently navigate this vastness to retrieve the most pertinent code segments given a natural language description? 
In the domain of software development, code search has become an essential pursuit for developers. Frequently, they dedicate significant time to combing through existing codebases in search of fragments that align with their needs. The aim of code search is to uncover code snippets within repositories that reflect users' intentions, often articulated in natural language. The proliferation of extensive code libraries, exemplified by platforms like GitHub and StackOverflow, has introduced a formidable challenge: efficiently retrieving semantically equivalent code from a vast array of possibilities \cite{deepcs,oppandcha,survey,tang2023app}.

In the past, coding was an isolated pursuit centered on translating logic into machine-readable instructions. Yet, modern development's collaborative landscape champions code reuse and modularity. Efficiently harnessing existing code is now essential, underscoring the need for a sophisticated code search mechanism. Such a mechanism must surpass syntax matching, comprehending the intricate semantics and intent in both code and queries. Early strategies, based on traditional information retrieval, relied on keyword matches \cite{portfolio, codehow, sourcerer}, lacking the nuance to untangle deep semantics or fathom natural language subtleties, often yielding suboptimal outcomes. The rise of deep learning and natural language processing heralded a transformative phase in code search \cite{deepcs, UNIF, TabCS, cdcs, CARLCS-CNN, CSRS, TranCS}, shifting towards encoding code and queries into dense semantic spaces to bridge the gap between abstract requirements and tangible implementations. This evolution is punctuated by notable landmarks: Information Retrieval Paradigms: Early code retrieval leaned on conventional methods, transforming queries and code using algorithms like TF-IDF \cite{portfolio, codehow, sourcerer}. Deep Learning Inroads: The resurgence of neural networks and deep learning shifted the landscape, encoding code and queries into dense vectors to grasp semantic subtleties. Pre-trained Model Epoch: Recent research embraced pre-trained models such as CodeBERT, CodeRetriever, and CoCoSoDa \cite{codebert, coderetriever, cocosoda}, harnessing extensive datasets and intricate training to bridge the divide between natural language and code.

While each research avenue has indubitably advanced the domain, the aspiration for an optimal code retrieval system remains unfulfilled. Each method, while groundbreaking in its own right, encapsulates inherent limitations. Furthermore, the intricate interplay between natural language descriptions and code is rife with latent relationships and differences that extant methodologies might not fully capture.

The realm of mathematics frequently unveils insights and tools that hold potential for addressing intricate challenges in diverse fields. In the context of code retrieval, one such mathematical concept, hyperbolic geometry, emerges as a promising contender. Unlike traditional Euclidean spaces, hyperbolic spaces excel at depicting hierarchical structures, which often underlie the relationship between code and its corresponding natural language description. In this context, our study poses a fundamental question: Can the distinctive attributes of hyperbolic spaces be harnessed to construct a more potent and semantically conscious code retrieval system? In this paper, we embark on a journey to address this inquiry. We introduce "Hyperbolic Code QA Matching" or \tool{}, a pioneering approach that reimagines the very essence of code retrieval. By seamlessly integrating the inherent qualities of hyperbolic spaces with cutting-edge embedding techniques, we aspire to encapsulate the intrinsic hierarchies and relationships inherent in the code-description interplay. We contend that this methodology, underpinned by the mathematical precision of hyperbolic geometry, has the potential to bridge the semantic gap between natural language queries and code with unparalleled efficiency.

Our contributions can be summarized as follows:
\begin{itemize}
    \item \textbf{Novel Hyperbolic Architecture}: We propose "Hyperbolic Code QA Matching" (\tool{}), an innovative approach that leverages Hyperbolic space to establish relationships between code fragments and queries, simplifying complex interaction mechanisms.
    \item \textbf{QA Framework Redefinition}: We redefine code retrieval as a question-answering framework, processing triple matches through a static BERT embedding layer to create initial embeddings.
    \item \textbf{Enhanced Efficacy}: Compared to existing solutions, \tool{} showcases superior performance, achieving 3.5\%-4\% average improvement against leading code retrieval methods.
\end{itemize}

\section{Approach}
\label{sec:approach}
\begin{figure}[!ht]
    \centering
    \includegraphics[width=\linewidth]{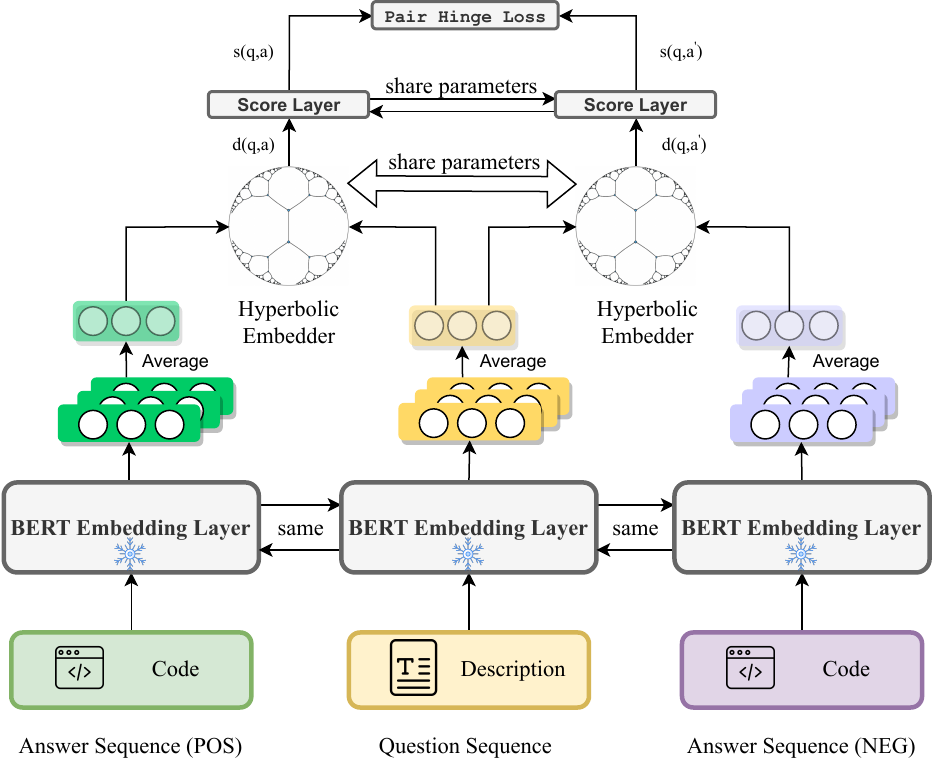}
    \caption{Architecture of \tool.}
    \label{fig:architecture}
\end{figure}

As illustrated in Figure~\ref{fig:architecture}, each description is paired with a correct code segment (positive answer) and an erroneous one (negative answer). Drawing inspiration from the natural language processing domain, we posit that the relationship between a description and its associated code mirrors the question-answer (QA) dynamic prevalent in NLP. Specifically, while the description elucidates the problem, the code delineates the solution to that problem. In this context, our primary objective is to maximize the margin between the scores of the correct QA pair and the negative QA pair, ensuring that the system can robustly differentiate between accurate and inaccurate solutions based on the given description.

\paragraph{Transformation: From Code Retrieval to QA Pair Matching}
In the code retrieval process, given a description, the objective is to identify and validate the presence of any pertinent code within the top N retrieved codes. To streamline this, we transform the code retrieval task into a QA pair matching paradigm: for a given description, we pair it with an accurate code (designated as positive) and a randomly selected inaccurate code (designated as negative). The primary training objective is to optimize the model to widen the gap between the scores of the accurate QA pair and the erroneous QA pair. During the testing phase, our refined hyperbolic model is employed to embed both codes and descriptions. Subsequently, for a presented description, the system evaluates the presence of the appropriate code among the top N retrieved codes.

\subsection{BERT Embedding Layer}
To adeptly comprehend the relationship between descriptions and code, it's imperative to translate textual sequences into their corresponding numerical representations. Our architecture processes three distinct sequences: the question (denoted as \( q \)), the accurate answer (symbolized as \( a \)), and a randomly chosen incorrect answer (referred to as \( a^{\prime} \)). Each of these sequences contains \( M \) words, where \( M_{q} \) and \( M_{a} \) represent the predetermined maximum sequence lengths for questions and answers, respectively.

While numerous methodologies~\cite{bafna2016document,pennington2014glove,church2017word2vec,joulin2016fasttext} exist for the transformation of text into vector representations, the optimal selection of an embedding technique is paramount. This choice directly impacts the fidelity with which the vectors capture textual nuances. Contrary to widely-adopted models like Word2Vec~\cite{mikolov2013efficient}, which assigns a static representation to each word irrespective of its surrounding context, BERT~\cite{devlin2019bert} offers a more nuanced approach. Specifically, BERT yields word vectors that dynamically adjust based on the context provided by adjacent words while in our work, we choose static BERT.

Owing to these merits, we employ BERT~\cite{devlin2019bert} as our foundational embedding model for both descriptions and tokenized code segments. Furthermore, in order to make our model's model faster, the parameters of BERT are frozen. The iteration we utilize is a pre-trained expansive model, comprising 24 layers and an embedding dimensionality of 1,024, fine-tuned on cased English literature. Once the text sequences are embedded in this vector space, it enables us to execute various numerical operations on them, such as determining textual similarity or computing correlation metrics.

\subsection{Task-Specific Word Representation via Projection}
To derive a word representation tailored to our task, we employ a projection layer. This layer is conceptualized as a singular neural network layer, impacting each word present in the three sequences.

\begin{equation}
    \begin{aligned}
    x &= \sigma\left(\mathbf{W}_{p} z + b_{p}\right) \\
    \text{subject to} \quad \mathbf{W}_{p} &\in \mathbb{R}^{d \times n}, \quad z \in \mathbb{R}^{n}, \quad x \in \mathbb{R}^{d}
    \end{aligned}
\end{equation}

where \( \sigma \) is a non-linear function, such as the rectified linear unit (ReLU). The outcome of this layer consists of a set of \( d \)-dimensional embeddings corresponding to each sequence, namely the question, the positive answer, and the negative answer. Crucially, the parameters intrinsic to this projection layer are consistently shared across the question and its associated answer.

\subsection{Deriving QA Representations}
To extract representations for questions and answers, we straightforwardly aggregate all word embeddings within the sequence.

\begin{equation}
y^{*} = \sum_{i=1}^{M_{*}} x_{i}^{*}
\end{equation}

In this equation, \( * \) encompasses \( \{q, a, a^{\prime}\} \). \( M \) denotes the preset maximum sequence length (pertinent to both question and answer), while \( x_{1}, x_{2}, \ldots, x_{M} \) are the \( d \)-dimensional sequence embeddings. Furthermore, we normalize the question and answer embeddings to fit within the unit sphere before progressing to subsequent layers, ensuring \( \|y^{*}\| \leq 1 \). This is achieved through \( y^{*} = \frac{y^{*}}{\|y^{*}\|} \) whenever \( \|y^{*}\| > 1 \). Emphasizing, this normalization of QA embeddings to the unit sphere is imperative for the optimal functionality of \tool.

\subsection{Embedding Interactions within a Hyperbolic Riemannian Framework for QA Pairs}
In the realm of neural ranking, the choice of interaction function between representations of questions and answers serves as a defining attribute. Within the scope of our research, we predominantly employ the hyperbolic distance function~${ }^{1}$ to elucidate the intricate relationships embedded within questions and answers. Explicitly, let's consider \(\mathcal{B}^{d}\) as the open \(d\)-dimensional unit ball, defined as \(\left\{ x \in \mathbb{R}^{d} \mid \|x\| < 1 \right\}\). Our model is conceptualized within the Riemannian manifold \(\left(\mathcal{B}^{d}, g_{x}\right)\) and is endowed with a specific Riemannian metric tensor, which can be expressed as:

\begin{equation}
    \begin{split}
    g_{x} &= \left(\frac{2}{1-\|x\|^{2}}\right)^{2} g^{E} \\
    \textit{s.t.}\quad g^{E} &\text{ is the Euclidean metric tensor}
    \end{split}
\end{equation}

Delving into the hyperbolic distance function that characterizes the interaction between the question and answer, it can be delineated as:

\begin{equation}
    \begin{split}
    d(q, a) &= \operatorname{arcosh}\left(1 + 2 \frac{\|q-a\|^{2}}{\left(1-\|q\|^{2}\right)\left(1-\|a\|^{2}\right)}\right) \\
    \textit{s.t.}\quad q, a &\in \mathbb{R}^{d}
    \end{split}
\end{equation}

The term "arcosh" is synonymous with the inverse hyperbolic cosine function, represented as \(\operatorname{arcosh} x = \ln \left(x + \sqrt{x^{2} - 1}\right)\). Notably, the value of \(d(q, a)\) exhibits a nuanced variation predicated on the spatial positioning of \(q\) and \(a\). This fluidity fosters the organic discovery of latent hierarchies. Given this configuration, an exponential surge in distance is observed as the vector's norm approaches unity. This phenomenon results in the encapsulation of inherent hierarchies within the QA embeddings through the vector's norm. From a geometric vantage point, the origin is visualized as a tree's root, proliferating expansively towards the periphery of the hyperbolic ball. The innate ability of the hyperbolic distance to discern hierarchies is elucidated both graphically and qualitatively in subsequent segments.

\subsubsection{Gradient Computation}
Among the various hyperbolic geometric models, the Poincarè hyperbolic distance stands out due to its differentiability. Given this, the partial derivative with respect to \(\theta\) is:

\begin{equation}
\begin{split}
\frac{\partial d(\theta, x)}{\partial \theta} &= \frac{4}{\beta \sqrt{\gamma^{2}-1}}\left(\frac{\|x\|^{2} - 2\langle\theta, x\rangle + 1}{\alpha^{2}} \theta - \frac{x}{\alpha}\right) \\
\textit{s.t.} \quad \alpha &= 1 - \|\theta\|^{2}, \\
\beta &= 1 - \|x\|^{2}, \\
\gamma &= 1 + \frac{2}{\alpha \beta}\|\theta-x\|^{2}.
\end{split}
\end{equation}
While various hyperbolic geometric models are available, such as the Beltrami-Klein and Hyperboloid models, our preference is the Poincarè ball/disk due to its differentiation simplicity and absence of constraints~\cite{nickel2017poincare}.

\subsection{Hyperbolic Distance-Based Similarity Computation}
In the intricate architecture of our model, the hyperbolic distance's transformation through a linear layer forms a pivotal step. This step ensures that the abstract spatial relationships in the hyperbolic space are mapped to values that can be utilized effectively in subsequent layers. The transformation is represented as:

\begin{equation}
    \begin{split}
    s(q, a) &= w_{f} d(q, a) + b_{f} \\
    \textit{s.t.}\quad w_{f} &\in \mathbb{R}^{1}, \\
    b_{f} &\in \mathbb{R}^{1}
    \end{split}
\end{equation}

The parameters, \( w_{f} \) and \( b_{f} \), are scalar components that govern this transformation, adjusting the scale and bias respectively. Their significance is underscored by empirical evidence: this layer has been chosen after a rigorous evaluation process, which considered various alternatives and their performance metrics.

\subsection{Optimization Techniques and Learning Paradigm}
The realm of optimization in neural architectures, especially those operating in non-Euclidean spaces, is vast and intricate. Within the \tool framework, the optimization strategy leans heavily on a pairwise ranking loss, aligning perfectly with the metric-centric nature of the model.

\subsubsection{Incorporation of Pairwise Hinge Loss}
To ensure the model discerns correct answers from incorrect ones effectively, it is trained to minimize a pairwise hinge loss. This loss function is articulated as:

\begin{equation}
    \begin{split}
    L &= \sum_{(q, a) \in \Delta_{q}} \sum_{\left(q, a^{\prime}\right) \notin \Delta_{q}} \max \left(0, s(q, a) + \lambda - s\left(q, a^{\prime}\right)\right) \\
    \textit{s.t.}\quad \Delta_{q} &\text{ is the set of all QA pairs for question } q
    \end{split}
\end{equation}

The incorporation of the pairwise hinge loss is not arbitrary; its selection is rooted in empirical results, demonstrating its superior performance in similar scenarios.

\subsubsection{Riemannian Optimization}
Navigating the landscape of hyperbolic space presents unique challenges, especially when it comes to gradient-based optimization. Recognizing this, our model utilizes Riemannian optimization techniques:

\begin{equation}
    \begin{split}
    \theta_{t+1} &= \mathfrak{R}_{\theta_{t}}\left(-\eta \nabla_{R} \ell\left(\theta_{t}\right)\right) \\
    \textit{s.t.}\quad \mathfrak{R}_{\theta_{t}} &\text{ denotes a retraction to } \mathcal{B} \text{ at } \theta
    \end{split}
\end{equation}

The Riemannian gradient has a close relationship with its Euclidean counterpart, which offers computational advantages:

\begin{equation}
    \begin{split}
    \nabla_{R} &= \frac{\left(1-\left\|\theta_{t}\right\|^{2}\right)^{2}}{4} \nabla_{E} \\
    \textit{s.t.}\quad \nabla_{E} &\text{ is the Euclidean gradient}
    \end{split}
\end{equation}

Owing to the complexity and nuances of working in hyperbolic space, we steer readers seeking a deeper understanding towards references~\cite{bonnabel2013stochastic,nickel2017poincare}. In the implementation phase, the power of TensorFlow's gradient computation is harnessed, albeit with necessary transformations as detailed above.

\subsection{Evaluative Retrieval during Testing with \textsc{\tool}}
\label{subsec:evaluative_retrieval}

\begin{figure}[H]
    \centering
    \includegraphics[width=\linewidth]{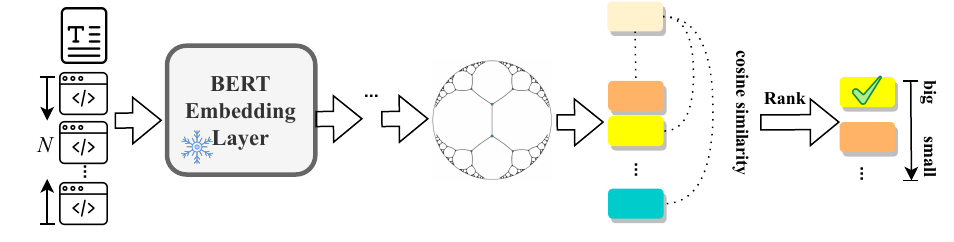}
    \caption{Test Stage}
    \label{fig:test}
\end{figure}

As shown in Figure~\ref{fig:test}, during the evaluative phase, the proficiency of the trained \tool{} comes to the fore, allowing for the discernment of well-matched question-answer pairs. Given a description, denoted as \( d \), and assuming the availability of \( N \) code snippets for retrieval, embeddings for \( d \) and the corresponding \( N \) code vectors are derived as per Equation~\ref{eq:tool}.

\begin{equation}
    \begin{split}
    C &= (\textbf{c}_1, \textbf{c}_2, \dots, \textbf{c}_N) \\
    \text{s.t.} \quad \textbf{c}_i &= \tool{} (c_i)
    \end{split}
    \label{eq:tool}
\end{equation}

Subsequent to obtaining the set \( C \) of embedded codes, it is ranked based on their relevance to the description \( d \). The primary evaluative criterion is the position of the ground truth code within this ranked set; specifically, we assess whether the actual code associated with \( d \) appears within the top \( N \) entries of \( C \).

\section{Experimental Design}\label{sec:exp}
In this section, we present our experimental setup, metrics, baselines, and research questions.

\subsection{Dataset}

\begin{figure}[H]
    \centering
    \includegraphics[width=\linewidth]{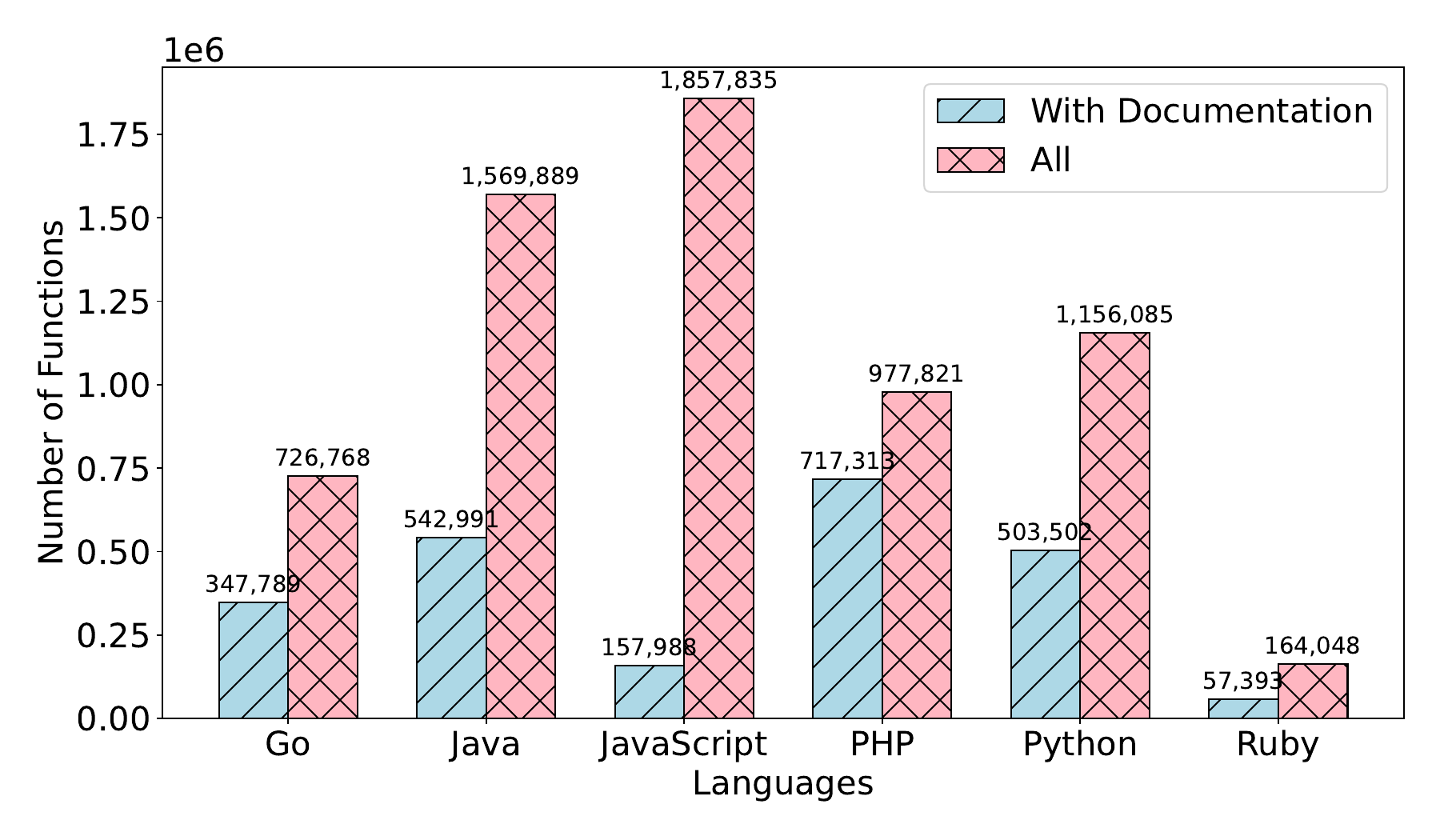}
    \caption{Dataset Size Statistics by Language}
    \label{fig:dataset}
\end{figure}

As depicted in Figure~\ref{fig:dataset}, the CodeSearchNet serves as a pivotal benchmark in the domain of code searching. Comprising over 2 million code snippets sourced from GitHub, this dataset spans six distinct programming languages: Go (726,768 snippets), Java (1,569,889 snippets), JavaScript (1,857,835 snippets), PHP (977,821 snippets), Python (1,156,085 snippets), and Ruby (164,048 snippets). The primary objective of CodeSearchNet is to facilitate developers in efficiently locating the requisite code. Furthermore, it catalyzes advancements in research areas such as natural language processing and code search methodologies.

It is imperative to note that the values presented in Figure~\ref{fig:dataset} also denote the quantity of positive pairs. Specifically, the number of snippets with documentation for each language are as follows: Go (347,789), Java (542,991), JavaScript (157,988), PHP (717,313), Python (503,502), and Ruby (57,393). In the context of our research, we employ a stochastic approach to select code not aligned with the ground truth to form a negative pair. Consequently, our dataset structure manifests as \texttt{<}positive code, description, negative code\texttt{>}, maintaining the dimensions elucidated in Figure~\ref{fig:dataset}.

\section{Evaluation Metrics}

In assessing the effectiveness of our proposed methodology, we adopt a series of metrics, which have been consistently recognized in prevailing research \cite{gu2018deep,du2021single,wan2019multi, he2020momentum}. Specifically, we employ the mean reciprocal rank (MRR) complemented by top-k recall, represented as \(R@k\) where \(k \in \{1,5,10\}\). The MRR metric furnishes a nuanced evaluation by ascertaining the average of the inverse ranks of the relevant code snippets corresponding to a designated set of queries, \(Q\). In contrast, \(R@k\) offers an aggregate metric by determining the proportion of queries wherein the associated code snippets are encompassed within the top-k entities of the resultant list. 

\begin{equation}
MRR = \frac{1}{|Q|} \sum_{i=1}^{|Q|} \frac{1}{\text{Rank}_i}
\end{equation}

\begin{equation}
R@k = \frac{1}{|Q|} \sum_{i=1}^{|Q|} \delta(\text{Rank}_i \leq k)
\end{equation}

In this context, \(\text{Rank}_i\) denotes the ranking of the code snippet that is paired with the \(i\)-th query within the resultant list. The function \(\delta\) serves as an indicator, producing a value of 1 if \(\text{Rank}_i \leq k\) and 0 otherwise.

\subsubsection{State-of-the-art}

\noindent
\textbf{CodeBERT~\cite{feng2020codebert}}:  Developed using the Transformer-based architecture and trained with a hybrid objective incorporating replaced token detection, CodeBERT efficiently leverages both bimodal data (NL-PL pairs) and unimodal data.

% \noindent
% \textbf{GraphCodeBERT~\cite{guo2020graphcodebert}}: While TwinRNN offers a commendable benchmark in the realm of patch detection, GraphSPD, presents an alternative perspective and methodology. 

% \noindent
% \textbf{GraphCodeBERT~\cite{guo2020graphcodebert}}: While TwinRNN offers a commendable benchmark in the realm of patch detection, GraphSPD, presents an alternative perspective and methodology. 

\noindent
\textbf{CodeRetriever~\cite{li2022coderetriever}}: CodeRetriever incorporates two contrastive learning schemes: unimodal contrastive learning, which employs an unsupervised approach to build semantically-related code pairs based on documentation and function names, and bimodal contrastive learning, which utilizes documentation and inline comments to form code-text pairs.

% \noindent
% \textbf{UniXcoder~\cite{guo2022unixcoder}}: UniXcoder is a unified cross-modal pre-trained model tailored for programming languages, enhancing code representation through mask attention matrices and cross-modal content like ASTs and code comments. The model introduces a unique approach to convert the tree-structured AST into a sequential format while retaining its structural essence.

\noindent
\textbf{CoCoSoDa~\cite{shi2023cocosoda}}: CoCoSoDa is a novel approach for code search, leveraging multimodal momentum contrastive learning and soft data augmentation to retrieve semantically relevant code snippets from natural language queries.

\subsection{Research Questions}
\begin{itemize}[leftmargin=*]
	\item {\bf RQ-1: How effective is {\em \tool{}} in code search?} 
	\item {\bf RQ-2: What is the impact of key design choices on the performance of {\em \tool{}}?}
        \item {\bf RQ-3: How do visualizations of \tool's representations differ between positive and negative pairs across programming languages?}
        \item {\bf RQ-4: How does {\em \tool{}} perform in real case?} 
\end{itemize}

% \subsection{Hyperparameter \& Environment}
% We take the only pretrained codeBERT model as the backbone. And the combined model of the backbone and adversarial learning component is used to initialize the vector representation. We use the Adam optimizer to optimize detector. We set the batch size to 64. We set the training to stop after 10 epochs. All experiments were performed on two 48G NVIDIA A6000 GPUs.
% \par
% There are three hyperparameters can be tuned in this experiment. We set the learning rate of classification task as  $lr_{cls} = 10^{-6}$ and the learning rate of dual contrastive learning task as $lr_{dcl} = 10^{-7}$. We study the code search performance of the model on $topK$ results, and set $K = \{1, 5\}$.
\section{Experiment Results}
\subsection{[RQ-1]: Effectiveness of \tool}
\label{subsec:rq1}
 \begin{figure*}[!h]
     \includegraphics[width=\linewidth]{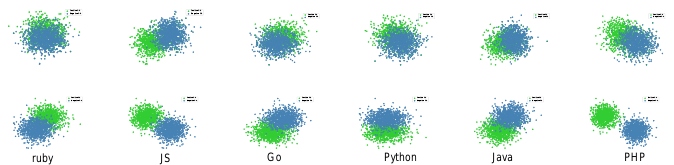}
     \caption{Visualization of compared models on positive and negative pairs from six programming languages. Green dots represent ``positive QA" and blue dots means ``negative QA". The first Row is visualized results of CodeRetriever and The second row is \tool{}'s.}
    \label{fig:intentiondetection}
 \end{figure*}
\noindent
\textbf{[Experiment Goal (RQ-1)]:}
In this study, our primary objective is to rigorously assess and benchmark the performance of our newly proposed \tool model, especially in the context of code search tasks. We have chosen the CodeSearchNet dataset for this evaluation due to its comprehensive coverage across six distinct programming languages. By pitting 
\tool against widely recognized and state-of-the-art benchmarks such as CodeBERT, CoCoSoDa, and CodeRetriever, we intend to draw informed comparisons and insights about its relative strengths and potential areas for improvement. Given the intricate nature of code retrieval and its implications for developer productivity, we emphasize the Mean Reciprocal Rank (MRR) as our primary metric of evaluation. Through this, we aspire to understand and quantify the tangible improvements and benefits that \tool might offer over existing models in the domain.

\noindent
\textbf{[Experiment Results (RQ-1)]:}

In our rigorous assessment of model performance on the CodeSearchNet dataset, spanning six diverse programming languages, we juxtaposed the capabilities of our proposed model, \tool, against three state-of-the-art benchmarks: CodeBERT, CoCoSoDa, and CodeRetriever. It's important to highlight that, in the absence of accessible results for CodeRetriever and considering the substantial resources required for its replication across the six programming languages, we undertook the task of reproducing CodeRetriever results for a comprehensive comparison.

The outcomes, as depicted in Table~\ref{tab:rq1}, are predominantly based on the MRR metric, chosen for its conciseness and the constraints imposed by space limitations. An inspection of the table reveals that 
\tool exhibits a commendable performance, consistently outstripping the MRR scores of its counterparts across all programming languages. Specifically, when contrasted with CoCoSoDa, which is one of the most competitive benchmarks, 
\tool demonstrates an improvement ranging from 3.5\% to 4\% across different languages, with an average enhancement of approximately 3.5\%. Such consistent and tangible increments in MRR values underscore the efficacy and robustness of our 
\tool model. Moreover, while CodeRetriever itself is a formidable contender, our \tool model surpasses it by marginal yet consistent increments, solidifying its position as a leading model in this domain.

\begin{table}[H]
\centering
\caption{Performance assessment of our methods is based on established metrics, with JS representing JavaScript. For the experiments, we maintained statistical significance at \(p<0.01\).}
\resizebox{1\linewidth}{!}{
\begin{tabular}{c|c|c|c|c|c|c|c}
\toprule
Model & Ruby & JS & Go & Python & Java & PHP & Avg. \\
\midrule
CodeBERT & 0.679 & 0.621 & 0.885 & 0.672 & 0.677 & 0.626 & 0.693 \\ \hline
CoCoSoDa & 0.818 & 0.764 & 0.921 & 0.757 & 0.763 & 0.703 & 0.788 \\ \hline
CodeRetrieval & 0.838 & 0.784 & 0.941 & 0.777 & 0.783 & 0.723 & 0.808 \\ \hline
\tool{} & \cellcolor{black!25}0.853 & \cellcolor{black!25}0.799 & \cellcolor{black!25}0.956 & \cellcolor{black!25}0.792 & \cellcolor{black!25}0.798 & \cellcolor{black!25}0.738 & \cellcolor{black!25}0.823 \\
\bottomrule
\end{tabular}}
\label{tab:rq1}
\end{table}

We also conduct recall experiment across all baselines. The table provides a detailed comparison of various models' performance using the Recall metric across six programming languages. Recall measures a model's ability to identify relevant items, and a higher value indicates better retrieval of pertinent items. CodeBERT, a renowned benchmark, showcases consistent recall scores across all languages, with Go and Java being particularly impressive. However, CoCoSoDa enhances upon CodeBERT, especially in Ruby and JavaScript, as evidenced by its higher R@1 metric. Interestingly, CodeRetriever surpasses CoCoSoDa across all metrics and languages, emphasizing its superior capability in retrieving a more extensive set of relevant items. Yet, the standout performer is \tool, which consistently outperforms all other models across every metric and programming language. This superior performance positions \tool as a potential leader in code search tasks, underlining its robustness and adaptability across various coding languages.

\begin{table}[!h]
    \centering
    \caption{Comparison on Recall Metric.}
    \resizebox{1\linewidth}{!}{
    \begin{tabular}{c|c|c|c|c|c|c|c}
    \toprule
    Models & Metric & Ruby & JavaScript & Go & Python & Java & PHP \\
    \midrule
    \multirow{3}{*}{CodeBERT} & R@1 & 0.583 & 0.514 & 0.837 & 0.574 & 0.580 & 0.520 \\
    \cline{2-8}
     & R@5 & 0.800 & 0.752 & 0.944 & 0.792 & 0.796 & 0.753 \\
    \cline{2-8}
     & R@10 & 0.853 & 0.814 & 0.962 & 0.850 & 0.852 & 0.814 \\
    \hline 
    \multirow{3}{*}{CoCoSoDa} & R@1 & 0.655 & 0.582 & 0.861 & 0.614 & 0.624 & 0.561 \\
    \cline{2-8}
     & R@5 & 0.875 & 0.806 & 0.962 & 0.834 & 0.843 & 0.798 \\
    \cline{2-8}
     & R@10 & 0.916 & 0.866 & 0.978 & 0.888 & 0.890 & 0.863 \\
    \hline 
    \multirow{3}{*}{CodeRetriever} & R@1 & 0.665 & 0.592 & 0.871 & 0.624 & 0.634 & 0.571 \\
    \cline{2-8}
     & R@5 & 0.885 & 0.816 & 0.972 & 0.844 & 0.853 & 0.808 \\
    \cline{2-8}
     & R@10 & 0.926 & 0.876 & 0.988 & 0.898 & 0.900 & 0.873 \\
    \hline 
    \multirow{3}{*}{\tool} & R@1 & \cellcolor{black!25}0.675 & \cellcolor{black!25}0.602 & \cellcolor{black!25}0.881 & \cellcolor{black!25}0.634 & \cellcolor{black!25}0.644 & \cellcolor{black!25}0.581 \\
    \cline{2-8}
     & R@5 & \cellcolor{black!25}0.895 & \cellcolor{black!25}0.826 & \cellcolor{black!25}0.982 & \cellcolor{black!25}0.854 & \cellcolor{black!25}0.863 & \cellcolor{black!25}0.818 \\
    \cline{2-8}
     & R@10 & \cellcolor{black!25}0.936 & \cellcolor{black!25}0.886 & \cellcolor{black!25}0.998 & \cellcolor{black!25}0.908 & \cellcolor{black!25}0.910 & \cellcolor{black!25}0.883 \\
    \bottomrule
    \end{tabular}}
    \label{tab:recall_result}
\end{table}

\find{{\bf \ding{45} Answer to RQ-1: }
The experimental results accentuate the potential of 
\tool in delivering superior performance in code search tasks across a gamut of programming languages. Experimental results indicate \tool{} outperforms previous works and achieve a 3.5-4\% improvement across languages in the term of MRR against the SOTA.}

\subsection{[RQ-2]: Ablation Study}
\label{subsec:rq2}

[\textbf{Experiment Goal}]: 
We perform an ablation study to investigate the effectiveness of each component in \tool. The major novelty of {\bf {\em \tool}} is the fact that it explicitly includes and processes: \textbf{{\em hp}} hyperbolic representation. In addition, we also evaluate the component of BERT.

\noindent
[\textbf{Experiment Design}]: We investigate the related contribution of {\em hp} and {\em bert} by building two variants of \linebreak \tool where we remove either  {\em hp}  (\ie denoted as \tool$_{hp-}$), or  {\em bert} (\ie denoted as \linebreak \tool$_{bert-}$). We evaluate the performance of these variants on the task of code search.

\begin{table}
\centering
\caption{Ablation Study}
\resizebox{1\linewidth}{!}{
\begin{tabular}{c|c|c|c|c|c|c|c}
\toprule
Model & Ruby & JS & Go & Python & Java & PHP & Avg. \\
\midrule
\tool{}$_{bert-}$ & 0.848 & 0.794 &0.951 & 0.787 & 0.793 & 0.733 & 0.818 \\
\tool{}$_{hp-}$ & 0.830 & 0.760 &0.920 & 0.752 & 0.758 & 0.700 & 0.787 \\
\tool{} & 0.853 & 0.799 &0.956 & 0.792 & 0.798 & 0.738 & 0.823 \\
\bottomrule
\end{tabular}}
\label{tab:ablation}
\end{table}

\noindent
\textbf{[Experiment Results (RQ-2)]:}

The performance dynamics of \textbf{\emph{\tool}} are intricately tied to its constituent components. To shed light on the contribution of each individual component, we conducted an ablation study. The results, as illustrated in Table \ref{tab:ablation}, pave the way for several enlightening insights. From the table, we observe that the absence of the hyperbolic representation component, denoted as \textbf{hp}, in \textbf{\emph{\tool}}$_{hp-}$ results in a noticeable performance degradation. The average score drops to \(0.787\), representing a decline of approximately \(4.38\%\) in comparison to the comprehensive \textbf{\emph{\tool}} model. On the other hand, when we omit the BERT component, leading to \textbf{\emph{\tool}}$_{bert-}$, the performance reduction is more modest. The average score settles at \(0.818\), a diminution of about \(0.61\%\). This indicates that while BERT plays a contributory role, it's the hyperbolic representation that stands out as the linchpin in enhancing the model's efficacy. In summary, the complete \textbf{\emph{\tool}} model, which amalgamates both BERT and hyperbolic representation, attains the best performance metrics. This underscores its robustness and adaptability in tackling code retrieval tasks across a spectrum of programming languages. The ablation study provides a roadmap for future research, highlighting areas of potential improvement and innovation.

\find{{\bf \ding{45} Answer to RQ-2: }
he ablation study of \textbf{\emph{\tool}} highlighted hyperbolic representation (\textbf{hp})'s crucial role. Its absence resulted in a substantial 4.38\% performance drop, while excluding the BERT component led to a mere 0.61\% decline. The integrated \textbf{\emph{\tool}} model, which combines both elements, showcased superior performance.
}

\subsection{[RQ-3]: Visualization of Learned Representation}
\label{subsec:rq3}

% Efficiency on Patch Intention Recognition
\noindent
\textbf{[Experiment Goal (RQ-3)]:} In our study, we train our model to maximize the margin between scores for correct and negative QA pairs, using visualization to understand code retrieval capabilities. Efficient QA visualization becomes a metric to assess model performance. We compare SOTA CodeRetriever with our \tool{} on six language-description QA pairs in this experiment.%In our work, the approach to train our model is to maximize the margin between the scores of the correct QA pair and the negative QA pair, which means visualizing positive QA and negative QA can help us to understand the ability of code retrieval of compared models. Thus, the efficiency of QA visualization can be used to measure if the code retrieval model is good or not. In this experiment, we choose the SOTA CodeRetrieval and \tool to do the visualization comparison on six language-description QA pairs. %1000 samples for each language

\noindent
\textbf{[Experiment Results (RQ-3)]:}

In our quest to comprehensively understand the discriminative capabilities of code retrieval models, we visualized the spatial distributions of positive and negative QA pairs. The underlying rationale behind this visualization is rooted in our training approach: our objective was to accentuate the margin between the scores of the correct QA pair and its negative counterpart. A clear demarcation between these two sets, when visualized, serves as a testament to the model's ability to efficiently retrieve relevant code snippets. A model's prowess in code retrieval can, thus, be gauged by the clarity and distinction it offers in such visualizations.

For a comparative perspective, we elected two models for this visualization task: the state-of-the-art model CodeRetriever and our proposed model, \tool. Our observations from the visualizations across all language-description QA pairs were illuminating. \tool demonstrated an evident superiority in distinguishing between positive QA and negative QA pairs. The distinct clusters formed by \tool were more segregated than those of CodeRetriever, underscoring its enhanced code retrieval capability. This clear visual distinction buttresses our assertion that \tool possesses a heightened ability to discern and retrieve relevant code based on given queries, outshining its contemporaries in this domain.

\find{{\bf \ding{45} Answer to RQ-3: }
Through visualizing positive and negative QA pairs, we evaluated code retrieval models' discriminative capabilities. While our training aimed to widen score margins between QA pairs, clarity in visualization proved the true test. Compared to the state-of-the-art CodeRetriever, \tool{} excelled in discerning QA pairs across programming languages, highlighting its advanced code retrieval proficiency.
}
\subsection{[RQ-4]: Case Study}
% We show some cases to demonstrate the effectiveness of our model \tool. For each case, we only show the result of our approach and the best baselines, which are CoCoSoDa, CodeBERT, and CodeRetrieval.

% \begin{figure}
%     \centering
%     \includegraphics[width=\linewidth]{pictures/casestudy.pdf}
%     \caption{Top-1 results by compared models.}
%     \label{fig:casestudy}
% \end{figure}

% As shown in Figure~\ref{fig:casestudy}, given a description ``a static method creating a Function from T to U using a given value", the ground truth is ``public static <T, U> Function<T, U> justFunction(U value)
% {
%        return new JustValue <T, U> (value); 
% }

To further elucidate the superiority of our model \textbf{\emph{\tool}}, we conducted a case study where we juxtaposed the predictions made by our model against those by the state-of-the-art baselines, namely CoCoSoDa, CodeBERT, and CodeRetriever. 

Given the prompt: ``a static method creating a Function from T to U using a given value", the ground truth code snippet is:

\begin{lstlisting}[language=Java]
public static <T, U> Function<T, U> justFunction(U value) {
       return new JustValue <T, U> (value); 
}
\end{lstlisting}

\textbf{\emph{\tool}} successfully predicts the ground truth while CodeRetriever's prediction:

\begin{lstlisting}[language=Java]
public static <X> Processor setupFunction(X xInput) {
    return new DifferentClass(xInput);
}
\end{lstlisting}

CoCoSoDa's prediction:

\begin{lstlisting}[language=Java]
public static <X, Y, Z> BiFunction<X, Z, Y> createComplexFunction(Y yParam, Z defaultParam) {
    ...
            return new AnotherFunctionClass<X, Y, Z>(yParam, zValue).apply(xValue, zValue);
        }
    };
}
\end{lstlisting}

In this instance, it is evident that \textbf{\emph{\tool}} offers a more accurate prediction in comparison to the baselines. Such cases underscore the robustness and precision of our model in understanding and generating code based on natural language descriptions.

\find{{\bf \ding{45} Answer to RQ-4: }
In our case study, \textbf{\emph{\tool}} precisely retrieved a description with a correct code, outperforming baselines like CoCoSoDa and CodeRetriever in accuracy and recall.
 }

\section{Related Work}
\subsection{Advancements in Code Representation}

Code representation learning is pivotal for numerous software engineering tasks like code summarization~\cite{iyer2016summarizing,leclair2019neural,shi2022evaluation}, code search~\cite{deepcs,li2020learning,haldar2020multi,du2021single}, and more. Particularly, code search aids significantly in software development and maintenance~\cite{singer2010examination,nie2016query}. While early methods~\cite{mcmillan2011portfolio}, \cite{lu2015query,lv2015codehow} leaned on lexical information retrieval, recent deep learning models embrace neural networks to enhance semantic code comprehension. Notable contributions include the use of sequential models~\cite{wan2019multi}, convolutional networks~\cite{li2020learning}, tree neural networks~\cite{wan2019multi}, graph models~\cite{wan2019multi,ling2021deep}, and transformers~\cite{du2021single,ocor}. Large-scale pre-trained models~\cite{graphcodebert,codebert,unixcoder}, \cite{niu2022spt} further enrich code semantics understanding, with exemplars like CodeBERT and GraphCodeBERT. Our method complements such pre-trained models, amplifying their efficacy.

\subsection{Neural Interactions in QA and Hyperbolic Potential}

While neural encoders like CNN or LSTM have proven their mettle in ranking models, recent focus gravitates towards the interaction layer. Initial models combined question and answer (QA) embeddings directly. Modern techniques, however, exploit similarity matrices, capturing nuanced word matches between QAs. Yet, these models, like AI-CNN or AP-BiLSTM~\cite{xu2017jointly,he2015multi,CARLCS-CNN,severyn2015learning,tay2017learning,tay2018cross,yu2014deep,zhang2017attentive,tay2018hyperbolic}, can be computationally demanding. Hyperbolic space offers an alternative, capturing hierarchical QA relationships efficiently.
\section{Conclusion}
In the evolving realm of software development, efficient code retrieval remains paramount. This study introduced the groundbreaking "Hyperbolic Code QA Matching" (\tool{}), marking a significant stride in code search methodologies. By ingeniously harnessing the hierarchical representation capabilities of hyperbolic spaces and synergizing them with advanced embedding techniques, we've offered a solution that transcends traditional lexical matching. Our approach delves deeper, capturing the intrinsic semantic relationships between natural language descriptions and code. The empirical results underscore the superior efficacy of \tool{}, setting a new benchmark in code search tasks. As the vast expanse of open-source platforms continues to grow, tools like \tool{} will become indispensable, empowering developers to navigate information oceans with unparalleled precision.

\section{Acknowledgments}
This work is supported by the NATURAL project, which has received funding from the European Research Council (ERC) under the European Union’s Horizon 2020 research and innovation programme (grant No. 949014).

\bibliography{aaai24}
\end{document}